
\magnification=\magstep1
\overfullrule=0pt
\def\lb{\lbrack}
\def\rb{\rbrack}
\def\q#1{\lb#1\rb}
\def\mn{\medskip\smallskip\noindent}
\def\sn{\smallskip\noindent}
\def\bn{\bigskip\noindent}
\font\extra=cmss10 scaled \magstep0 \font\extras=cmss10 scaled 750

\setbox1 = \hbox{{{\extra R}}}
\setbox2 = \hbox{{{\extra I}}}
\setbox3 = \hbox{{{\extra C}}}

\def\C{{{\extra C}}\hskip-\wd3\hskip2.5 true pt{{\extra I}}\hskip-\wd2
\hskip-2.5 true pt\hskip\wd3}
\def\Complex{\hbox{{\extra\C}}}
\def\One{{{\extra 1}}\hskip-\wd3\hskip0.5 true pt{{\extra 1}}\hskip-\wd2
\hskip-0.5 true pt\hskip\wd3}
\def\id{\hbox{{\extra\One}}}
\setbox4=\hbox{{{\extra Z}}}
\setbox5=\hbox{{{\extras Z}}}
\setbox6=\hbox{{{\extras z}}}
\def\Z{{{\extra Z}}\hskip-\wd4\hskip 2.5 true pt{{\extra Z}}}
\def\z{{{\extras Z}}\hskip-\wd5\hskip 2 true pt{{\extras Z}}}

\def\Zed{\hbox{{\extra\Z}}}
\def\zed{\hbox{{\extras\z}}}

\def\mod{\phantom{l} {\rm mod} \phantom{l}}
\def\vac{\mid \!v \rangle\, }
\def\avac{\langle v \! \mid\, }
\def\state#1{\mid \! #1 \rangle\, }
\def\rrangle{\rangle \hskip -1pt \rangle}

\def\astate#1{\langle #1 \! \mid\, }
\def\pstate#1{\Vert #1 \rrangle\, }


\def\chapsubtitle#1{\leftline{\bf #1}
\vskip-10pt
\line{\hrulefill}}
\def\vl{\hskip 1pt \vrule}
\def\vt{\tilde{v}}
\def\vact{\mid \!\vt \rangle\, }
\def\avact{\langle \vt \! \mid\, }
\def\normvac{\langle v \! \mid \!v \rangle\, }
\def\normvact{\langle \vt \! \mid \!\vt \rangle\, }
\def\ab{\bar{\alpha}}
\def\a{\alpha}
\def\si{\sigma}
\def\Ga{\Gamma}
\def\om{\omega}
\def\la{\lambda}
\def\vphi{\varphi}
\def\rt{\tilde{r}}

\def\H{{\cal H}}

\def\O{{\cal O}}
\def\GS#1{\state{{\rm GS};#1}}
\def\bsin#1{\sin{\left(#1\right)}}
\def\bcos#1{\cos{\left(#1\right)}}
\def\cab{{\cal C}}
\def\Rab{{\cal R}}
%
\def\secA{1}
\def\secB{2}
\def\secC{3}
\def\secD{4}
\def\secE{5}
\def\secF{6}
\def\appA{A}
\def\hkn{1}
\def\gehri{2}
\def\dogra{3}
\def\daviesA{4}
\def\daviesB{5}
\def\onsager{6}
\def\albertiniA{7}
\def\albertiniB{8}
\def\yang{9}
\def\baxter{10}
\def\scm{11}
\def\hela{12}
\def\tang{13}
\def\weA{14}
\def\dkcoy{15}
\def\gehkra{16}
\def\gehlenph{17}
\def\alcaraz{18}
\def\fateev{19}
\def\lykyanov{20}
\def\zam{21}
\def\mccoyadv{22}
\def\chrihen{23}
\def\gehritell{24}
\def\baym{25}
\def\hoeger{26}
\def\weB{27}
\def\elitzur{28}
\def\horn{29}
\def\albcoy{30}
\font\HUGE=cmbx12 scaled \magstep4
\font\Huge=cmbx10 scaled \magstep4
\font\Large=cmr12 scaled \magstep3

\font\large=cmr17 scaled \magstep0
\font\ZZfont=cmss17 scaled \magstep0
\setbox9=\hbox{{{\ZZfont Z}}}
\def\ZZ{{{\ZZfont Z}}\hskip-\wd9\hskip 4 true pt{{\ZZfont Z}}}
\def\ZZed{\hbox{{\ZZfont\ZZ}}}
%
%
\nopagenumbers
\pageno = 0
\centerline{\HUGE Universit\"at Bonn}
\vskip 10pt
\centerline{\Huge Physikalisches Institut}
\vskip 1.5cm
\centerline{\Large Low-Temperature Expansions}
\vskip 6pt
\centerline{\Large \phantom{g} and \phantom{g}}
\vskip 6pt
\centerline{\Large Correlation Functions of the
                   $\ZZed_{\hbox{\rm 3}}$-Chiral Potts Model}
\vskip 0.7cm
\centerline{by}
\vskip 0.5cm
\centerline{\large N.S.\ Han${\hbox{\rm 1} \atop \ }$)
               and A.\ Honecker${\hbox{\rm 2} \atop \ }$) }
\vskip 1.2cm
\centerline{\bf Abstract}
\vskip 11pt
\noindent
Using perturbative methods we derive new results for the
spectrum and correlation functions of the general $\Zed_3$-chiral Potts
quantum chain in the massive low-temperature phase. Explicit calculations
of the ground state energy and the first excitations in the
zero momentum sector give excellent approximations and confirm
the general statement that the spectrum in the low-temperature
phase of general $\Zed_n$-spin quantum chains is identical to
one in the high-temperature phase where the r\^ole of charge and
boundary conditions are interchanged. Using a perturbative
expansion of the ground state for the $\Zed_3$ model
we are able to gain some insight in correlation functions.
We argue that they might be oscillating
and give estimates for the oscillation length as well as the
correlation length.
\vfill
\line{\hbox to 5cm{\hrulefill} \hfill}
\line{${}^{1})$ {\bf Permanent address:} Department of Theoretical Physics,
         Hanoi State University, \hfill}
\line{\phantom{${}^{1})$ {\bf Permanent address:}} \hskip 1pt
         P.O.\ Box 600, Bo Ho, Hanoi, Vietnam \hfill}
\vskip 1.5cm
\settabs \+&  \hskip 110mm & \phantom{XXXXXXXXXXX} & \cr
\+ & Post address:                       & BONN-HE-93-13   & \cr
\+ & Nu{\ss}allee 12                     & hep-th/9304083  & \cr
\+ & W-5300 Bonn 1                       & Bonn University & \cr
\+ & Germany                             & April 1993      & \cr
\+ & ${}^{2})$ e-mail:                   & ISSN-0172-8733  & \cr
\+ & unp06b@ibm.rhrz.uni-bonn.de         &  \              & \cr
\eject
\pageno=1
\footline{\hss\tenrm\folio\hss}
\chapsubtitle{\secA.\ Introduction}
\mn
The self-dual $\zed_3$-chiral Potts model was introduced by
Howes, Kadanoff and den Nijs $\q{\hkn}$
and studied using e.g.\ fermionization and approximative methods,
in particular perturbation expansions. One remarkable result of
the perturbation expansions was that for special values of the
parameters the first translationally invariant excitation is
linear in the inverse temperature $\la$ for special values of the parameters.
v.\ Gehlen and Rittenberg then realized $\q{\gehri}$ that this model
is integrable for these special values of the parameters because it satisfies
the Dolan-Grady integrability condition $\q{\dogra}$ which
is equivalent $\q{\daviesA} \q{\daviesB}$ to Onsager's
algebra $\q{\onsager}$. They
also generalized this `superintegrable' chiral Potts model
to general $\Zed_n$-spin $n$ $\q{\gehri}$.
Afterwards, it attracted much attention because it can be related
to a classical model that satisfies a generalized Yang-Baxter equation
with Boltzmann weights defined on higher genus Riemannian surfaces
$\q{\albertiniA - \scm}$.
However, even then perturbative methods lead to important new results
$\q{\hela}\q{\tang}$. One example is a conjecture for the exact
form of the order parameters in general superintegrable
$\Zed_n$-chiral Potts chains $\q{\tang}$.
\sn
Recently, a particle interpretation of the
momentum zero sectors in the high-temperature phase of all
$\Zed_n$-chiral Potts models at general values of the parameters has been
proposed $\q{\weA}$ and a quasi-particle spectrum has been derived
for the superintegrable $\Zed_3$-chiral Potts model $\q{\dkcoy}$.
Furthermore, a scaling exponent for the wave vector in the
low-temperature phase of the $\Zed_3$-chiral Potts model has
been calculated in $\q{\gehkra}\q{\gehlenph}$ from level crossings
in the ground
state. This motivated us to perform the perturbative calculations
reported in this paper. On the one hand the excitation spectrum
in the low-temperature phase of general $\Zed_n$-chiral Potts
quantum chains is not completely understood. On the other hand
very little is known about correlation functions. Perturbation
expansions enable us to gain more insight in the spectrum of
the low-temperature phase and shed some light on correlation
functions.
\medskip
A general $\Zed_n$-spin quantum chain with $N$ sites is defined by
the Hamiltonian:
$$H^{(n)}_N = - \sum_{j=1}^N \sum_{k=1}^{n-1} \tilde{\la} \ab_k \si_j^k
              + \a_k \Ga_j^k \Ga_{j+1}^{n-k}.
\eqno({\rm \secA.1})$$
For $n=2$ (\secA.1) is just the well known Ising model. In this
case the operators $\si_j$ and $\Ga_j$ are the Pauli spin matrices
$\si_x$ and $\si_z$ acting in a vector space $\Complex^2$ located
at site $j$. For general $n$ one may think of the operators $\si_j$
and $\Ga_j$ as generalizations of the Pauli spin matrices -- see
(\secB.3) below. In this paper we identify the $N+1$st site with
the 1st site, i.e.\ we use toroidal boundary conditions.
\medskip
The Hamiltonians (\secA.1) contain $2n-1$ parameters. The temperature like
parameter $\tilde{\la}$ will be chosen real while the coupling
constants $\ab_k$ and $\a_k$ will be generally complex. $H^{(n)}_N$
is hermitian iff $\ab_k = \ab_{n-k}^{*}$ and $\a_k = \a_{n-k}^{*}$.
\mn
In this paper we will parametrize the constants $\ab_k$, $\a_k$ in (\secA.1)
by two angles $\phi$, $\vphi$, fixing their dependence on $k$:
$$\a_k = {e^{i \phi ({2 k \over n}-1)} \over \sin{\pi k \over n} } \ ,
\qquad
\ab_k = {e^{i \vphi ({2 k \over n}-1)} \over \sin{\pi k \over n} }.
    \eqno{(\rm \secA.2)}$$
(\secA.2) is called the general `chiral Potts model'.
This parametrization is convenient because it can easily be specialized
to various models. Setting $\phi = \vphi =0$
yields models with a second order phase transition at $\lambda=1$ that
can be described by a parafermionic conformal field theory in the limit
$N \to \infty$ $\q{\alcaraz}$. These so-called Fateev-Zamolodchikov-models
lead to extended conformal algebras ${\cal WA}_{n-1}$ where the simple
fields have conformal dimension $2, \ldots, n$ $\q{\fateev} \q{\lykyanov}$.
The spectrum of the Hamiltonian (\secA.1) with $\phi = \vphi = 0$
can be described by the first unitary minimal
model of the algebra ${\cal WA}_{n-1}$. For $n=3$ it
coincides with the three-states Potts model and the symmetry algebra
is Zamolodchikov's well known spin-three extended conformal algebra
$\q{\zam}$ at $c={4 \over 5}$.
\mn
For $\phi = \vphi = {\pi \over 2}$ (\secA.2) specializes to the
`superintegrable' $\Zed_n$-chiral Potts model which exhibits
remarkable integrability properties $\q{\gehri}$:
At $\phi = \vphi = {\pi \over 2}$ the
Hamiltonians (\secA.1) satisfy the Dolan-Grady
integrability condition $\q{\dogra}$.
\mn
Albertini et al.\ $\q{\albertiniA - \yang}$ have shown that the
Hamiltonian (\secA.1) can be obtained for more general values of the angles
$\phi$, $\vphi$ as the $\tau$-continuum limit of an integrable
classical statistical model if one imposes the constraint
$$\tilde{\la} \cos \vphi = \cos \phi
.    \eqno{(\rm \secA.3)}$$
$H^{(n)}_N$ with the choices (\secA.2), (\secA.3) is in general no more
self-dual. However, if we choose $\phi=\vphi$ in  (\secA.2) $H^{(n)}_N$
is self-dual. Sometimes (\secA.3) is implied when referring to the chiral
Potts model but we prefer to
call (\secA.1) with (\secA.2) the general chiral Potts model.
\medskip
The $\Zed_3$ version of (\secA.1) is known to have four phases
$\q{\albertiniB} \q{\mccoyadv} \q{\gehlenph}$: Two massive and two
massless phases. One of the massive phases is ordered and the other
massive phase is disordered. The low-temperature phase (small $\tilde{\la}$)
that we are going to study in this paper is the ordered massive phase.
At $\phi = \vphi = {\pi \over 2}$ it appears in the range $0 \le \tilde{\la}
\le 0.901292\ldots$ $\q{\albertiniB} \q{\mccoyadv}$.
\medskip
In the next section we will review some well known facts about the
Hamiltonian (\secA.1). Then, in section {\secC} we will evaluate the ground
state energy and the first excitations perturbatively
for $n=3$ and zero momentum, but arbitrary $\phi$, $\vphi$. The results
remind us of a general duality statement that we discuss in section \secD.
Finally, section {\secE} is devoted to a study of correlation functions
of the $\Zed_3$-chain.
\bn
\chapsubtitle{\secB.\ Preliminaries}
\mn
In this section we summarize well known facts about $\Zed_n$-quantum
spin chains and introduce notations that will be useful later on.
For a more detailed, recent review see e.g.\ $\q{\chrihen}$.
\mn
First, we give a precise definition of the operators $\Ga_j$
and $\si_j$.
$\si_j$ and $\Ga_j$ freely generate a finite dimensional
associative algebra by the following relations ($1 \le j,l \le N$):
$$\eqalign{
\si_j \si_l &= \si_l \si_j \ , \qquad
\si_j \Ga_l = \Ga_l \si_j \om^{\delta_{j,l}}_{} \ , \cr
\Ga_j \Ga_l &= \Ga_l \Ga_j \ , \qquad
\si_j^n = \Ga_j^n = \id \cr
} \eqno{(\rm \secB.1)}$$
where $\om$ is the $n$th root of unity
$\om = e^{2 \pi i \over n}_{}$. In this paper we will only consider
boundary conditions of type $\Gamma_{N+1} = \om^{-R} \Gamma_1$,
$R \in \Zed_n$ for the Hamiltonian (\secA.1). We will mainly focus on
periodic boundary conditions $\Ga_{N+1} = \Ga_1$.
\medskip
The algebra (\secB.1) can be conveniently represented in
$\otimes^N \Complex^n$.
In this space we can choose the following basis if we label
the standard basis of $\Complex^n$ by $\{ e_0, \ldots, e_{n-1} \}$:
$$\state{i_1 \ldots i_N} := e_{i_1} \otimes \ldots \otimes e_{i_N}
\qquad 0 \le i_j \le n-1.   \eqno{(\rm \secB.2)}$$
Usually one considers a special representation $r$ of the algebra
(\secB.1) -- a definition can be found in appendix \appA.
However, for low-temperature expansions of (\secA.1)
it is more convenient to consider a different representation $\rt$:
$$\eqalign{
\rt(\Ga_j) \state{i_1 \ldots i_j \ldots i_N}
            &= \om^{i_j} \state{i_1 \ldots i_j \ldots i_N} \cr
\rt(\si_j) \state{i_1 \ldots i_j \ldots i_N}
            &= \state{i_1 \ldots (i_j-1 \mod n) \ldots i_N}. \cr
}         \eqno{(\rm \secB.3)}$$
\medskip
$H^{(n)}_N$ commutes with the $\Zed_n$ charge operator
$\hat{Q} = \prod_{j=1}^N \si_j$, thus has $n$ charge sectors.
The eigenvalues of $\hat{Q}$ have the form $\om^Q$ with
$Q \in \{0, \ldots, n-1 \}$. We will refer to the number $Q$ as the `charge'.
\medskip
$H^{(n)}_N$ also commutes with the translation operator $T_N$.
The eigenvalues of $T_N$ are $N$th roots of unity. We label them by
$e^{i P}$ and call $P$ the `momentum'. For a chain of length $N$ one
has $P \in \{0, {2 \pi \over N}, \ldots, {2 \pi (N-1) \over N} \}$.
The eigenstates $\pstate{i_1 \ldots i_{N}}_P^{}$ with momentum $P$
can be obtained from $\state{i_1 \ldots i_{N} }$ by finite Fourier
transformation. $T_N$ acts for $R=0$ on the states (\secB.2) as
$$\rt(T_N) \state{i_1 i_2 \dots i_{N}} = \state{i_2  \ldots i_N i_1}.
     \eqno{(\rm \secB.4)}$$
In the case $R=0$ the eigenstates with momentum $P$ are given by
$$\pstate{i_1 \ldots i_{N}}_P^{} := {1 \over \sqrt{{\cal N}}}
     \sum_{x=0}^{N-1}  e^{i P x} \rt(T_N)^{-x} \state{i_1 \ldots i_{N} }.
\eqno{(\rm \secB.5)}$$
${\cal N}$ is a suitable normalization constant. If the state
$\state{i_1 \ldots i_{N} }$ has no symmetry, one has ${\cal N} = N$.
This will apply to all cases below where we need (\secB.5).
For a definition of the momentum eigenstates (\secB.5) in case
of boundary conditions $R \ne 0$ see e.g.\ $\q{\gehritell}\q{\chrihen}$
\bn
\chapsubtitle{\secC.\ Spectrum in the low-temperature phase}
\mn
In this section we calculate the ground state energy
and the lowest excitations in the low-temperature phase using
perturbative expansions according to $\q{\baym}$ around
$\tilde{\la} = 0$ of the Hamiltonian (\secA.1).
We restrict ourselves to periodic boundary conditions $R=0$.
\sn
In $\q{\hkn}$ perturbation series were presented for the
disorder operator (or magnetization $m$) and the first
energy gap in the momentum zero sector of the superintegrable
$\Zed_3$-chiral Potts model leading to exact conjectures for
both of them. After the superintegrable chiral Potts model
had been generalized to general $n$ $\q{\gehri}$ perturbation
series for the ground state energy, energy gap in the momentum
zero sector, magnetization and susceptibility of this superintegrable
$\Zed_n$-chiral Potts model were presented in $\q{\hela}$.
At the same time elaborate expansions of the ground state energy
and some excitations of the superintegrable chiral Potts model
for $n \in \{ 3, 4, 5 \}$ and in particular perturbation series
for the order parameters with general $n$ were calculated in
$\q{\tang}$. First perturbative results for the energy gaps
at more general values of the angles $\phi$, $\vphi$ were
obtained in $\q{\weA}$ where second order high-temperature
expansions for the translationally invariant energy gaps in
each charge sector of the general self-dual $\Zed_3$- and
$\Zed_4$-chiral Potts models as well as a first order expansion
for the dispersion relations for general $n$ was presented.
\sn
In this section we restrict once again to the $\Zed_3$-version
of the chiral Potts model (\secA.1) but impose no restrictions on
the angles $\phi$, $\vphi$. We present low-temperature
expansions for the ground state energy and in particular the first
translationally invariant energy gaps that up to now have not
been treated by perturbative methods because of high degeneracies.
\medskip
The normalization of the Hamiltonian (\secA.1) is chosen such
that expansions around zero temperature $\tilde{\la} = 0$
are possible. If one wants to calculate expansions around infinite
temperature one usually normalizes the Hamiltonian
$\hat{H}^{(n)}_N = {1 \over \tilde{\la}} H^{(n)}_N$,
sets $\la = \tilde{\la}^{-1}$ and performs expansions around
$\la = 0$.
\mn
In each charge sector $Q$ of the low-temperature phase there is
one unique ground state. For arbitrary $n$ it is given by:
$$\GS{Q} := {1 \over \sqrt{n}} \sum_{l=0}^{n-1} \om^{l \cdot Q}
  \state{l \ldots l}     \eqno{(\rm \secC.1)}$$
provided that $-{\pi \over 2} \leq \phi \leq {\pi \over 2}$.
For $n=3$ (\secC.1) is the ground state if $-\pi \leq \phi \leq \pi$ and for
$n=4$ (\secC.1) is the ground state for
$-{5 \pi \over 6} \leq \phi \leq {5 \pi \over 6}$.
The excited states are more complicated and highly degenerate. The
space of the first excitation is spanned by those states which
have precisely two blocks of different spins.
Furthermore, the values of the spins in these two blocks must
have difference one. For fixed
$P$, $Q$ and $-{\pi \over 2} \leq \phi \leq {\pi \over 2}$
we can choose the following basis for the space of the first excitation:
$$\state{a_k^Q} := {1 \over \sqrt{n}} \sum_{l=0}^{n-1} \om^{l \cdot Q}
  \pstate{\underbrace{(l+1 \mod n) \ldots (l+1 \mod n)}_{k \phantom{l}
                                     {\rm times}} l \ldots l}_P^{}.
  \eqno{(\rm \secC.2)}$$
In order to perform explicit calculations we now specialize to
$n=3$ with $P = 0$. The constant contribution in $\tilde{\la}$ to the
ground state energy $E_{\GS{Q}}$ and the first gap $\Delta E_{Q,1}$
can be calculated easily:
$$E_{\GS{Q}}^{(0)} = -N {4 \over \sqrt{3}} \bcos{{\phi \over 3}} \ ,
\qquad\qquad
\Delta E_{Q,1}^{(0)} = 4 \sqrt{3} \bcos{{\phi \over 3}}.
      \eqno({\rm \secC.3})$$
As far as the ground state is concerned, we notice from the explicit form
of the potential in (\secA.1) that the first order in the perturbation
expansion vanishes. The next order that has to be calculated is the second
order $E_{\GS{Q}}^{(2)}$. One applies the potential
$V = \sum_{j,k} \ab_k \si_j^k$ to the ground state:
$$\rt(V) \GS{Q} = -{2 \sqrt{N} \over \sqrt{3}}
    \left(e^{{i \vphi \over 3}} \state{a_1^Q}
       +e^{-{i \vphi \over 3}} \om^Q \state{a_{N-1}^Q} \right).
\eqno({\rm \secC.4})$$
Using $\Delta E_{Q,1}^{(0)} = E_{\state{a_k^Q} }^{(0)} - E_{\GS{Q}}^{(0)}$
one obtains the result
$$\eqalign{
E_{\GS{Q}}^{(2)} &= \sum_{\state{a} \neq \GS{Q}}
   {\mid \astate{a} \rt(V) \GS{Q} \mid^2 \over
     E_{\GS{Q}}^{(0)} - E_{\state{a} }^{(0)}} \cr
&= - {\mid \astate{a_1^Q} \rt(V) \GS{Q} \mid^2 +
      \mid \astate{a_{N-1}^Q} \rt(V) \GS{Q} \mid^2
      \over \Delta E_{Q,1}^{(0)}} \cr
&= - {2 N \over 3 \sqrt{3} \bcos{\phi \over 3} }. \cr
}      \eqno({\rm \secC.5})$$
Set $\cab := \bcos{{\phi \over 3}}$,
$\Rab := 1 - 4 \cab^2$. Then, calculating higher orders in the same manner,
one arrives at:
$$\eqalign{
N e_0 := E_{\GS{Q}}(N) =
             -N {\Biggl(} & {4 \over \sqrt{3}} \cab
             + {2 \over 3 \sqrt{3} \cab } \tilde{\la}^2
             + {\cos{\vphi} \over
             9 \sqrt{3} \cab^2} \tilde{\la}^3
             - {\sqrt{3} \over 81 \cab } \left \{
             {1 \over 2 \cab^2 }
             +{4 \over \Rab }
             \right \} \tilde{\la}^4 \cr
           & - {\sqrt{3} \cos{\vphi} \over 81 \cab^2}
              \left \{ {3 \over 4 \cab^2 } + {4 \over \Rab }
             \right \} \tilde{\la}^5
                   {\Biggr)}    + \O(\tilde{\la}^6).  \cr
}  \eqno{(\rm \secC.6)}$$
for sufficiently large $N$. For $\vphi = \phi = {\pi \over 2}$
(\secC.6) reproduces the result of $\q{\hela}$.
The orders $e_0^{(k)}$ of the free energy per site $e_0$
are independent of $N$ if $N > k$.
This is a general feature of the ground state energy for spin
quantum chains with nearest neighbour interaction.
\mn
Note that the expansion in powers of $\tilde{\la}$ (\secC.6)
of $e_0$ does not depend on the charge sector for large $N$.
{}From the explicit calculations we see
that this is a general result: The  order $\tilde{\la}^k e_0^{(k)}$
of the free energy does not depend on the charge if $N > k$.
However, ground state level crossings have been observed in
$\q{\gehkra}\q{\gehlenph}$. Indeed, for short chain length $N$
high orders $\tilde{\la}^k$ ($k \ge N$) do depend on the charge $Q$.
Thus, at fixed $\tilde{\la}$ in the massive phase level crossings
in the ground state do occur although the $Q$-dependent term
of the ground state energy $e_0$ decreases fast in magnitude with
increasing $N$. The presence of such level crosssings is a hint
for oscillating correlation functions and, in fact, the critical
exponent of the wave vector can be calculated from them $\q{\hoeger}
\q{\gehkra}\q{\gehlenph}$. For the superintegrable case
$\phi = \vphi = {\pi \over 2}$ we explicitly determined
level crossings in the ground state using 9th resp.\ 10th order
expansions for $3 \le N \le 7$ sites. Our results for the
temperatures $\tilde{\la}$ where the gap vanishes are in good
agreement with the values presented in $\q{\gehlenph}$ (Table 3)
and show in particular that there are no crossings besides
the one presented in $\q{\gehlenph}$. However, we have argued
that for larger $N$ we would need even higher orders for the
study of level crossings which goes beyond current computer
power. Thus, numerical methods provide a much better tool
for the study of level crossings $\q{\gehkra}\q{\gehlenph}$.
\mn
Finally, we should mention that the approximation (\secC.6) is
excellent for the whole massive low-temperature phase up to
its boundary near $\tilde{\la} = 1$ already for moderately long
chains. For example at the phase transition
$\tilde{\la} = 1$ of the parity conserving Potts case
$\phi = \vphi =0$ one observes only a deviation of $0.8\%$
between (\secC.6) and the exact result for $N=12$ sites.
In the superintegrable case $\phi = \vphi = {\pi \over 2}$
at $\tilde{\la} = 0.9$ (close to the boundary of the phase)
the deviation of (\secC.6) at $N=12$ from the exact result
is also as small as $1.3\%$.
Thus, the perturbation expansion (\secC.6) yields a surprisingly
good approximation in the complete massive low-temperature phase.
Even at the boundary of the phase the deviation is smaller
than $2\%$ for $N>10$.
\mn
In principle, one can derive a critical exponent $\alpha$ for the specific
heat ${{\rm d}^2 e_0 \over {\rm d}\tilde{\la}^2}$ from a perturbation
expansion of the ground state energy $e_0$. However, (\secC.6) leads only
to a third order expansion for the specific heat and on the
superintegrable line even the third order vanishes. Although
the approximation of (\secC.6) to the ground state energy $e_0$
itself is so good, one certainly needs higher orders
for accurate estimates of the critical exponent of the specific heat.
Consequently, $\alpha$ has been estimated using a 13th order
expansion of $e_0$ in $\q{\hkn}$ for the self-dual case $\phi = \vphi = 0$
and for the superintegrable case $\phi = \vphi = {\pi \over 2}$
the Ising-like form of the eigenvalues has been exploited to calculate
even higher orders of $e_0$ in $\q{\tang}$. The results in $\q{\hkn}\q{\tang}$
indicate $\alpha = {1 \over 3}$ independent of the angles $\phi$, $\vphi$.
\bigskip
The calculation of the smallest gap $\Delta E_{Q,1}$ is more difficult.
Let $q$ be the projector onto the space spanned by the states
$\state{a_k^Q}$ (\secC.2). In this space, the potential acts as follows:
$$\eqalign{
q \rt(V) \state{a_1^Q} &= -{2 \over \sqrt{3}}
   (2 e^{{i \vphi \over 3}} \state{a_2^Q}
   + e^{{i \vphi \over 3}} \om^Q \state{a_{N-1}^Q}) \cr
q \rt(V) \state{a_{k}^Q} &= -{2 \over \sqrt{3}}
   (2 e^{{i \vphi \over 3}} \state{a_{k+1}^Q}
   + 2 e^{-{i \vphi \over 3}} \state{a_{k-1}^Q} )
   \qquad 1 < k < N-1 \cr
q \rt(V) \state{a_{N-1}^Q} &= -{2 \over \sqrt{3}}
   (2 e^{-{i \vphi \over 3}} \state{a_{N-2}^Q}
   + e^{-{i \vphi \over 3}} \om^{2 Q} \state{a_1^Q}). \cr
}  \eqno{(\rm \secC.7)}$$
In the limit $N \to \infty$ the lowest perturbative eigenvector
converges to
$${1 \over \sqrt{N-1}} \sum_{k=1}^{N-1} \state{a_k^Q}.
  \eqno{(\rm \secC.8)}$$
Using (\secC.7) and (\secC.8) we can calculate that for $N \to \infty$
and $\phi < {\pi \over 2}$
$$\eqalign{
\lim_{N \to \infty} \Delta E_{Q,1} = &
   4 \sqrt{3} \cab
   - \tilde{\la} {8 \over \sqrt{3}} \bcos{{\vphi \over 3}} \cr
  &+ \tilde{\la}^2 \left\{ {8\cab \left(1+\bcos{{2\vphi \over 3}}\right)
                 \over 3 \sqrt{3} (\Rab + 2)}
   + {4 \left(2 - \bcos{{2 \vphi \over 3}} \right) \over 3 \sqrt{3} \cab }
        \right\}
   + \O(\tilde{\la}^3). \cr
}  \eqno{(\rm \secC.9)}$$
Comparing (\secC.9) with the corresponding high-temperature expansion
$\q{\weA}\q{\weB}$ shows that up to the order calculated it coincides with
$\Delta E_{1,0}(\la) + \Delta E_{2,0}(\la) $ at the dual point
$\la = \tilde{\la}$ $\phi$, $\vphi$ with interchanged. In fact,
we can also estimate the error we have made in calculating (\secC.9). The
error we are making when replacing the true eigenvector by (\secC.8) is
of the magnitude $N^{-1}$. This behaviour is preserved when the
potential $V$ and the resolvent $g$ are applied.
Thus, we obtain in all orders starting with the first one a deviation which
is of magnitude $N^{-2}$. This further supports the identification of
(\secC.9) with the dual of a two-particle state.
Of course, (\secC.9) holds only for $\phi < {\pi \over 2}$.
\sn
For $\phi \ge {\pi \over 2}$, the states (\secC.2) are not the first excited
states any more. Now we have to consider the following states:
$$\eqalign{
{1 \over \sqrt{n}} {\bigg (} &
   \pstate{1 \ldots 1 \ 2 \ldots 2 \ldots 0 \ldots 0 }_P^{}
   + \om^Q \pstate{2 \ldots 2 \ldots 1 \ldots 1 }_P^{} \cr
 & + \ldots
   + \om^{Q (n-1)} \pstate{0 \ldots 0 \ldots (n-1) \ldots (n-1)}_P^{}
          {\bigg )}.        \cr
}  \eqno{(\rm \secC.10)}$$
Going through the same steps as before we find:
$$\eqalign{
\lim_{N \to \infty} \Delta E_{Q,1} = &
   12 \bsin{{\pi - \phi \over 3}}
   - \tilde{\la} 4 \sqrt{3} \bcos{{\vphi \over 3}} \cr
  &- \tilde{\la}^2 {2 \over \sqrt{3} } \left\{
      {\bcos{{2 \vphi \over 3}} - 2 \over \cab }
   +  {\bcos{{2 \vphi \over 3}} + 1 \over \bcos{{\pi - \phi \over 3}} }
      \right\}
   + \O(\tilde{\la}^3),
\qquad \phi \ge {\pi \over 2}. \cr
}  \eqno{(\rm \secC.11)}$$
(\secC.11) coincides with $3 \Delta E_{1,0}(\la)$ at the dual point
$\la = \tilde{\la}$ with interchanged $\phi$, $\vphi$ in the corresponding
high-temperature expansion $\q{\weA}\q{\weB}$ up to the order calculated.
\sn
For $\phi= {\pi  \over 2}$ the states (\secC.2) and (\secC.10) are
degenerate. However, for large $N$
the dominant contribution comes from the states (\secC.10)
such that (\secC.11) is valid for $\phi = {\pi \over 2}$ as well.
At $\phi = \vphi = {\pi \over 2}$ this is in agreement with the exact
result of $\q{\baxter}$:
$$\lim_{N \to \infty} \Delta E_{Q,1} =
   6 (1 - \tilde{\la})
  \qquad {\rm for} \phantom{x} \phi = \vphi = {\pi \over 2}.
  \eqno{(\rm \secC.12)}$$
\bn
\chapsubtitle{\secD.\ Duality of spectra}
\mn
The results in the previous section remind us of some well known
results about duality: The spectra in the low-temperature phase
at $\tilde{\la}$ are dual to those in the high-temperature phase
at $\la=\tilde{\la}$ if we interchange $\a_k$ and $\ab_k$.
This statement for $\Zed_n$ quantum spin chains has been known
for a long time $\q{\elitzur} \q{\horn}$ and was also used
in $\q{\hkn}$. However, special attention has to be
paid to the boundary conditions when performing duality transformations.
It has been observed in $\q{\gehri}$ that the duality transformation
interchanges the r\^ole of the charge $Q$ and boundary conditions $R$.
In this section we reformulate the precise statement of duality
for the general $\Zed_n$-chiral Potts quantum chain (\secA.1)
and discuss its consequences. For completeness, a simple non-standard
proof is presented in appendix A.
\medskip
We denote the Hamiltonian (\secA.1) including the parameters
by $H_N^{(n)}(\tilde{\la},R^{lt},\ab_k^{lt},\a_k^{lt})$. If the
Hamiltonoperator is properly normalized for high-temperature
expansions ($\hat{H}(\la) := \la H(\la^{-1})$ ) it will be called
$\hat{H}_N^{(n)}(\la,R^{ht},\ab_k^{ht},\a_k^{ht})$, writing again
explicitly the corresponding parameters. In order
to be able to distinguish the parameters we have introduced an
upper index. Furthermore, abbreviate the space with charge $Q^{ht}$ in the
high-temperature phase by $\H^{Q^{ht}}$ and the eigenspace of $\rt(\hat{Q})$
to eigenvalue $\om^{Q^{lt}}$ by $\tilde{\H}^{Q^{lt}}$. Now we can formulate
the statement: $\hat{H}_N^{(n)}(\la,R^{ht},\ab_k^{ht},\a_k^{ht})$ restricted
to $\H^{Q^{ht}}$ and
$H_N^{(n)}(\tilde{\la},R^{lt},\ab_k^{lt},\a_k^{lt})$ restricted to
$\tilde{\H}^{Q^{lt}}$ have the same spectra if
$$\eqalign{
Q^{lt} &= R^{ht} \ , \qquad  R^{lt} = Q^{ht} \ , \cr
\ab_k^{lt} &= \a_k^{ht} \ , \qquad  \a_k^{lt} = \ab_k^{ht} \ , \qquad
\tilde{\la} =  \la. \cr
}  \eqno{(\rm \secD.1)}$$
The momentum decomposition can be applied alike in the high- and
low-temperature phase. Thus, the statement of duality
is also valid if we further restrict to eigenspaces with
momentum $P$.
\medskip
Note that the duality (\secD.1) preserves the condition (\secA.3).
Thus, each integrable chiral Potts model is dual to exactly one
integrable chiral Potts model. The integrable model is self-dual iff
$\la=1$ and $\phi = \vphi \in {\pi \over 2} \Zed$.
The point $\la=1$, $\phi = \vphi = 0$ exhibits conformal invariance
$\q{\alcaraz}$. However, this is not true for the other self-dual
points. For example the point $\la=1$, $\phi = \vphi = {\pi \over 2}$
is not conformally invariant although in its vicinity non-diagonal
conformal field theories have been used to derive correlation functions
$\q{\albcoy}$.
\medskip
{}From (\secD.1) we conclude that the
quasi-particle interpretation for the high-temperature
phase of the general $\Zed_n$-chiral Potts quantum chain $\q{\weB}$
can be pulled over to the low-temperature phase. The duality
transformation interchanges charge sector $Q$ and boundary conditions
$R$. Thus, the ground state of the high-temperature
phase is mapped to periodic boundary conditions $R=0$
in the low-temperature phase. However, the fundamental excitations
are mapped to different boundary conditions corresponding to
the charge sectors $R \in \{1, \ldots n-1\}$. Therefore we
observed only composite particle states in section \secC.
\medskip
It is well known that in the high-temperature phase the limit $N \to \infty$
of the energy eigenvalues is
independent of the boundary conditions whereas the charge is substantial.
The duality (\secD.1) implies that in the low-temperature phase this limit
does not depend on the charge, but the spectra are clearly different for
different boundary conditions. The independence from the charge has already
been observed in section {\secC} for some eigenvalues. One can also argue
directly that this degeneracy holds for the complete spectra, at least in the
range where the perturbation expansion converges.
\sn
In the perturbation expansion all orders can be organized with respect to
the energy eigenvalues at zero temperature. All energy eigenspaces
-- apart from the ground state -- have a dimension that grows at least
with $N$. We have seen in section {\secC} for some examples that at a
fixed order $k$ of the perturbation expansion only a finite number
$N_{E,k}$ of matrix elements of the potential $V^k$ in the eigenspace
to energy $E$ depend on the charge $Q$. This holds generally as we can see
from the proof of duality presented in appendix A (in particular (\appA.2)
and (\appA.3) ). Thus, the term for energy $E$ at order $k$ in the
perturbation expansion has a $Q$-dependent term that is at most of order
$({N_{E,k} \over N})^2$ as $N \to \infty$. This implies that the differences
between the charge sectors converge to zero in the large chain limit.
\sn
It remains to check that this is also true for the ground state.
The contribution of the ground state to a perturbation expansion
for any energy level other than for the ground state energy itself is
neglegible. The $k$th order of the ground state energy per site
$e_0^{(k)}$ is independent of the charge sector $Q$ if $k < N$. This is
easy to see because in the perturbation expansion application
of $k$ powers of the potential $V$ to the ground state is projected
back onto the ground state. Thus, all excitations that are created
must be annihilated again, or an excited state is proportional to
the ground state. The second case is only possible if an excitation
is carried around the boundary of the chain implying
$k \ge N$. On the other hand, the first possibility yields results
that are clearly independent of the charge sector. Finally, convergence of
the perturbation series for $e_0$ implies that the limit
$N \to \infty$ is independent of the charge sector $Q$.
\sn
In summary,
in the low-temperature regime all charge sectors are degenerate for
$N \to \infty$, at least if the perturbation expansion converges.
\bn
\chapsubtitle{\secE.\ Correlation functions in the low-temperature regime}
\mn
In this section we will apply methods explained in more detail in
$\q{\weB}$
to the correlation functions in the low-temperature phase of the
chiral $\Zed_3$ Potts quantum chain. Note that the duality argument
of section {\secD} applies only to the Hamiltonian and not to other operators.
Thus, quantities like e.g.\ correlation lengths may be different
in these two phases. In fact, current knowledge about the correlation
functions is restricted to the conjectures presented in $\q{\mccoyadv}$
and the argumentation in favour of a non-vanishing wave vector presented
in $\q{\gehkra}\q{\gehlenph}$.
\sn
We study the correlation functions
$$C_{\Ga}(x) := {\avac \Ga_x^{+} \Ga_0 \vac  \over \normvac} \ , \qquad
C_{\si}(x) := {\avac \si_x^{+} \si_0 \vac  \over \normvac}
              - {\avac \si_x^{+} \vac \avac \si_0 \vac \over \normvac^2}.
    \eqno({\rm \secE.1})$$
using a perturbative expansion $\q{\baym}$ for the ground state $\vac$
from the state $\GS{Q}$ (\secC.1). The expansion of the ground state
in powers of $\tilde{\la}$ leads to an expansion of the correlation
functions in powers of $\tilde{\la}$:
$$C_{\Ga}(x) = \sum_{k=0}^{\infty} \tilde{\la}^k C_{\Ga}^{(k)}(x) \ , \qquad
C_{\si}(x) = \sum_{k=0}^{\infty} \tilde{\la}^k C_{\si}^{(k)}(x).
    \eqno({\rm \secE.2})$$
In the definition (\secE.1) it is legitimate
to omit the one point functions for the operator
$\Ga$ because they are zero due to charge conservation.
\medskip
Below, we will first give the final results for general angles
$\phi$, $\vphi$. It turns out that the result is too complicated
to infer the general form of the correlation functions
(\secE.1). Thus, we then specialize to the superintegrable case
$\phi = \vphi = {\pi \over 2}$ and calculate even higher orders.
By looking for a good fit we try to guess the structure of the
correlation functions. With this experience we turn back
to the general case and discuss how the correlation functions
should change for general $\phi$, $\vphi$.
\medskip
In order to save space we present only the final results
for the correlation functions. For $C_{\Ga}(x)$ one obtains,
using again the abbreviations $\cab = \bcos{{\phi \over 3}}$,
$\Rab = 1 - 4 \cab^2$:
$$\eqalign{
C_{\Ga}^{(0)}(x) =& 1 \ , \qquad \qquad \qquad \qquad \qquad
C_{\Ga}^{(1)}(x) = 0 \ , \cr
C_{\Ga}^{(2)}(x) =& {1 \over 6 \cab^2} \left\{ \delta_{x,0} - 1 \right\}
          \ , \qquad \qquad
C_{\Ga}^{(3)}(x) = {\cos{\vphi} \over 18 \cab^3}
           \left\{ \delta_{x,0} - 1 \right\} \ , \cr
C_{\Ga}^{(4)}(x) =& \cr
       {1 \over 27 \cab^2} {\Biggl\{} &
           (1 - \delta_{x,0}) \left( { 2 ( 1 - 16 \cab^2) \over 3 \Rab^2}
                  + {1 \over 16 \cab^2} \right)
                   + \delta_{x,1} \left(
   {1 + 2 \cab^2 - 3 \ i \ \bsin{{2 \phi \over 3}} \over 3 \Rab^2}
                    + {5 \over 16 \cab^2} \right)
                      {\Biggr\}}. \cr
}   \eqno({\rm \secE.3})$$
The first orders of the correlation function $C_{\si}(x)$ read
as follows:
$$\eqalign{
C_{\si}^{(0)}(x) =& \delta_{x,0} \ , \qquad \qquad \qquad \qquad \qquad
                                     \qquad \quad \hskip 5pt
C_{\si}^{(1)}(x) = 0 \ , \cr
C_{\si}^{(2)}(x) =& -{1 \over 9} {\Biggl\{} {\delta_{x,0} \over \cab^2}
                               + \delta_{x,1} \left(
                           {2 \over \Rab}
                         + {1 \over 2 \cab^2} \right) {\Biggr\} } \ , \quad
C_{\si}^{(3)}(x) = -{\cos{\vphi} \over 9 \cab } {\Biggl\{}
                           {\delta_{x,0}  \over 2 \cab^2}
                      + {\delta_{x,1} \over \Rab} {\Biggr\} } \ , \cr
C_{\si}^{(4)}(x) =&  \cr
   {1 \over 81 \cab^2 } {\Biggl\{} \delta_{x,0} & \left(
                              {7 \over 16 \cab^2 }
                             +{8 \over  \Rab} \right)
                         +  \delta_{x,1} \left( {15 \over 16 \cab^2 }
                            + {29 \over 12 \Rab }
                            - {16  \cab^4 \over \Rab^3} \right)
                         + \delta_{x,2} \left(
                            - {1 \over 8 \cab^2 }
                            + {7 + 44 \cab^2 \over 12 \Rab^2}
                        \right) {\Biggr\} }. \cr
}   \eqno({\rm \secE.4})$$
Note that the correlation functions (\secE.3) and (\secE.4) do not
depend on the charge sector.
\medskip
(\secE.3) suggests that the correlation function $C_{\Ga}(x)$ tends to a
non-zero constant for large distances $x$ -- in contrast to the
correlation functions in the high-temperature phase $\q{\weB}$ so
that that the low-temperature phase is ordered over long ranges.
However, beyond this general conclusion, it is difficult to guess from
(\secE.3) or (\secE.4) what might be the behaviour even for small $x$.
Thus, we set $\phi = \vphi = {\pi \over 2}$, calculate four
further orders and obtain
$$\eqalign{
C_{\Ga}^{(0)}(x) =& 1 \ , \qquad \qquad \qquad \qquad
C_{\Ga}^{(1)}(x) =
C_{\Ga}^{(3)}(x) =
C_{\Ga}^{(5)}(x) =
C_{\Ga}^{(7)}(x) = 0 \ , \cr
C_{\Ga}^{(2)}(x) =& {2 \over 9} \left( \delta_{x,0} - 1 \right) \ , \cr
C_{\Ga}^{(4)}(x) =& {7 \over 81} \left( \delta_{x,0} - 1 \right)
                   + \delta_{x,1} {5 - i \sqrt{3} \over 162} \ , \cr
C_{\Ga}^{(6)}(x) =& {1 \over 6561} \left\{
                        336 \left(\delta_{x,0} - 1 \right)
                   + 160 \delta_{x,1} + 60 \delta_{x,2} \right\}
                   - i {\sqrt{3} \over 6561} \left\{
                     26 \delta_{x,1} + 20 \delta_{x,2} \right\} \ , \cr
C_{\Ga}^{(8)}(x) =& {1 \over 354294} \left\{
                        12600 \left(\delta_{x,0} - 1 \right)
                   + 6852 \delta_{x,1} + 3521 \delta_{x,2}
                   + 1225 \delta_{x,3} \right\} \cr
                  &- i {\sqrt{3} \over 354294} \left\{
                     960 \delta_{x,1} + 995 \delta_{x,2}
                   + 525 \delta_{x,3} \right\} \cr
}   \eqno({\rm \secE.5})$$
and
$$\eqalign{
C_{\si}^{(0)}(x) =& \delta_{x,0} \ , \qquad \qquad \qquad \qquad
C_{\si}^{(1)}(x) =
C_{\si}^{(3)}(x) =
C_{\si}^{(5)}(x) =
C_{\si}^{(7)}(x) = 0 \ , \cr
C_{\si}^{(2)}(x) =& {1 \over 27} \left\{ -4 \delta_{x,0}
                               + \delta_{x,1} \right\} \ , \cr
C_{\si}^{(4)}(x) =& {1 \over 729}\left\{
                        - 41 \delta_{x,0} + 14 \delta_{x,1} + 8 \delta_{x,2}
                    \right\} \ , \cr
C_{\si}^{(6)}(x) =& {1 \over 19683}\left\{
                - 586 \delta_{x,0} + 147 \delta_{x,1} + 126 \delta_{x,2}
                     + 80 \delta_{x,3} \right\} \ , \cr
C_{\si}^{(8)}(x) =& {1 \over 531441}\left\{
                - 9927 \delta_{x,0} + 2130 \delta_{x,1} + 1721 \delta_{x,2}
                   + \kappa \delta_{x,3}  + 910 \delta_{x,4} \right\} . \cr
}   \eqno({\rm \secE.6})$$
Unfortunately, determination of the constant $\kappa$ in $C_{\si}^{(8)}$
exceeded the numerical range of our special purpose computer algebra
system.
\sn
Up to the order calculated, $C_{\si}(x)$ is real for all values
of the parameters $\phi$, $\vphi$, $\tilde{\la}$.
$C_{\Ga}(x)$, in contrast, has a non-vanishing imaginary part.
By analogy to the high-temperature regime $\q{\weB}$ and from the results
in $\q{\gehkra}\q{\gehlenph}$ one might expect that $C_{\Ga}(x)$ is
oscillating. Indeed, the correlation functions in the superintegrable case
$\phi = \vphi = {\pi \over 2}$ can nicely be fitted by
$$\eqalignno{
C_{\si}(x) &= a \delta_{x,0} + b e^{-{x \over \xi_{\si}}} \ ,
                                             &({\rm \secE.7a})\cr
C_{\Ga}(x) &= m^2 +
      p e^{-\left({1 \over \xi_{\Ga}} + {2 \pi i \over L}\right) x}.
                                             &({\rm \secE.7b})\cr
}$$
For $\tilde{\la} \in \left\{ {1 \over 4}, {1 \over 2}, {3 \over 4} \right\}$
good fits to (\secE.5) and (\secE.6) using (\secE.7) are given by the values
in the following table:
\mn
\centerline{\vbox{
\hbox{
\vrule \hskip 1pt
\vbox{ \offinterlineskip
\def\tablespace{height2pt&\omit&\vl&\omit&&\omit&&\omit&\vl&\omit&&
                          \omit&&\omit&&\omit&&\omit&\cr}
\def\tablerule{ \tablespace
                \noalign{\hrule}
                \tablespace        }
\hrule
\halign{&\vrule#&
  \strut\hskip 2pt\hfil#\hfil\hskip 2pt\cr
\tablespace
\tablespace
&$\tilde{\la}$ &\vl& $\xi_{\si}$ && $a$       && $b$      &\vl&
            $\xi_{\Ga}$ && $m^2$       && $L$  &&
                      $\bsin{{2 \pi \over L}}$  && $p$
                          & \cr \tablespace \tablerule
& $0.25$       &\vl& $0.26(1)$   && $0.89(1)$ && $0.10(1)$ &\vl&
            $0.24(3)$   && $0.9857605$ && $30 \pm 18$ &&
                      $0.26(3)$                 && $0.011(3)$
                          & \cr \tablespace
& $0.50$       &\vl& $0.41(3)$   && $0.85(1)$ && $0.11(1)$ &\vl&
            $0.37(5)$   && $0.9381$    && $30 \pm 15$ &&
                      $0.27(1)$                 && $0.05(1)$
                          & \cr \tablespace
& $0.75$       &\vl& $0.59(9)$   && $0.75(2)$ && $0.14(2)$ &\vl&
            $0.50(7)$   && $0.832$     && $30 \pm 13$ &&
                      $0.27(1)$                 && $0.15(2)$
                          & \cr \tablespace
}
\hrule}\hskip 1pt \vrule}
\hbox{\quad Table 1: Parameters for the correlation functions
            (\secE.7) at $\phi = \vphi = {\pi \over 2}$ }}
}
\mn
First, we remark that the correlation lengths satisfy
$\xi_{\si} = \xi_{\Ga} =: \xi$ for all values
of $\tilde{\la}$ within the numerical accuracy. In fact, one
expects this equality because the correlation lengths should be
the inverses of some mass scale, and there is only one mass scale
in our problem because all three charge sectors are degenerate.
Furthermore, we observe that our data is
compatible with an oscillating correlation function for
the operator $\Ga$. The oscillation length $L$ (or wave vector)
is around $30$ sites in a major part of the low-temperature phase.
In $\q{\gehkra}\q{\gehlenph}$ it has been predicted that $L$
should diverge as $\tilde{\la}$ crosses the phase boundary
and approaches $\tilde{\la} = 1$ where the critical exponent
is expected to equal ${2 \over 3}$. Our results are compatible
with a divergent oscillation length at $\tilde{\la} = 1$
although due to the large errors we do not even see that
$L$ increases with $\tilde{\la}$.
\sn
We should mention that the linear approximation
$e^{-{2 \pi i \over L} x} \approx 1 - i x \bsin{{2 \pi \over L}}$ is as
good as (\secE.7b). However, (\secE.7b) seems to be a more natural form.
The fact that the relative error of the estimate for
$q$ is much smaller than that of $L$ just comes from the
fact that $L$ and $q$ are related by exponentiation.
\sn
Note that the short correlation lengths strongly damp the
correlation functions: At $x=7$ where we expect the first
zero of $C_{\Ga}(x)$ it has already decreased by at least six orders
of magnitude for $\tilde{\la} \le {3 \over 4}$. Closer to
$\tilde{\la} = 1$ the correlation length should increase but
so should the oscillation length as well. Thus, it will be very
difficult to obtain more precise results from approximative
arguments and an exact expression for $C_{\Ga}(x)$ is
probably needed in order to decide whether (\secE.7b) really
is the correct form and to determine the wave vector $L$
accurately.
\sn
Before we conclude the discussion of the correlation functions
for the superintegrable chiral Potts model, we mention that a conjecture
for the form of $C_{\Ga}(x)$ has been formulated in $\q{\mccoyadv}$:
$C_{\Ga}(x) = m^2 + \O ( e^{-{x \over \xi_{\Ga}}})$
where $m$ is the order parameter. Our result (\secE.7) is compatible
with this from. In $\q{\hkn}$ the conjecture
for the order parameter
$$m = {\avact \Ga_x \vact \over \normvact} =
      \left( 1 - \tilde{\la}^2 \right)^{1 \over 9}
   \eqno({\rm \secE.8})$$
has been formulated, but (\secE.8) has not been proven yet.
The constant term in (\secE.5) is in exact agreement with
(\secE.7) and (\secE.8) up to the order calculated,
such that we may assume that at least the constant term of $C_{\Ga}(x)$
is now known exactly.
\medskip
A few remarks on the choice of ground state (\secC.1) are
in place because  in $\q{\hkn} \q{\yang}$ (\secE.8) has actually been
derived considering an expectation value of the operator $\Ga_x$.
We have already pointed out that the one point functions of $\Ga_x$
vanish identically due to charge conservation if one uses the charge
eigenstates (\secC.1). However, if one uses instead non-charge
eigenstates like $\state{0 \ldots 0}$ for a perturbative expansion
of $\vact$ they do not vanish. Indeed, using an expansion for $\vact$
from $\state{0 \ldots 0}$ we once again verified equality of this
one point function with the order parameter $m$. If we redefine
$\tilde{C}_{\Ga}(x)$ by replacing $\vac$ by $\vact$ and
subtracting the contribution from the one point functions, this
is in fact the only change, i.e.\ $\tilde{C}_{\Ga}(x)=
C_{\Ga}(x)-m^2$. $C_{\si}(x)$ remains unchanged under this redefinition.
\medskip
For more general values of the angles $\phi$, $\vphi$ one
expects the correlation functions to be also of the form (\secE.7) --
of course with different values of the parameters. We can
see from the constant term $m^2$ of the correlation function $C_{\Ga}(x)$
(\secE.5) that it will not be of the form (\secE.8) for general
$\phi \ne {\pi \over 2} \ne \vphi$. In general, the coefficient
of $\tilde{\la}^3$ for the constant term does not vanish and
$m^2$ does not even have an expansion in powers of $\tilde{\la}^2$.
Among the powers that we have calculated for the general case
only the fourth order in (\secE.5) has a non-vanishing imaginary
part at $x=1$. Under the assumption that (\secE.7b) is the general
form we would expect the imaginary part at $x=1$ to be proportional
to $\bsin{{2 \pi \over L}}$ for very small temperatures $\tilde{\la}$.
Thus, we expect for very low temperatures $\tilde{\la}$ the relation
$L^{-1} \sim \phi$. On the one hand, this explains the conjectured
presence of a second length scale $L$ in addition to the correlation
length $\xi$. The oscillation length $L$ just comes from the chiral
angle $\phi$ and thus these two scales must be related to each other.
On the other hand, the oscillation (should it really be present)
will vanish smoothly as the parity conserving Potts
case $\phi = \vphi = 0$ is approached.
\bn
\chapsubtitle{\secF.\ Conclusion}
\mn
In this paper we discussed the low-temperature phase of the
$\Zed_3$-chiral Potts quantum chain.
We have perturbatively calculated the ground state energy and the
first gaps for $P=0$ at general chiral angles $\phi$, $\vphi$.
We explicitly observed duality
to the high-temperature phase and independence of the charge sector
$Q$. This demonstrates a general duality property stating
equality of spectra in the low- and high-temperature phase.
Thus, a quasi-particle interpretation for the high-temperature
phase of the general $\Zed_n$-chiral Potts quantum chain $\q{\weB}$
can be pulled over to the low-temperature phase. However, charge
$Q$ and boundary conditions $R$ are interchanged by the duality
transformation. In particular,
for periodic boundary conditions one sees only energy levels
above the ground state that correspond to composite particle
states. We also gave a general argument that in the infinite chain
length limit all charge sectors are degenerate.
\mn
We have further studied correlation functions for the
operators $\si$ and $\Ga$ in the low-temperature phase of the
$\Zed_3$-chiral Potts quantum chain. The correlation function
$C_{\Ga}(x)$ has a constant term $m^2$ indicating long range
order. Fitting (complex) exponential
functions to the perturbation expansions of the correlation
functions we estimated the correlation length $\xi$
for $\phi = \vphi = {\pi \over 2}$ and found agreement with
the prediction $\q{\gehkra} \q{\gehlenph}$
that the correlation function $C_{\Ga}$ oscillates.
A rough estimate for the oscillation length $L$ has also
been obtained. We argued that the oscillation length
should satisfy $L \sim \phi^{-1}$ for small temperatures.
In particular, the oscillation vanishes for vanishing chiral
angle $\phi$. These first results make an exact determination
of the correlation functions desirable.
\bn
\chapsubtitle{Acknowledgements}
\mn
It is a pleasure to thank H.\ Hinrichsen and G.v.\ Gehlen for numerous
valuable discussions.
\sn
One of us (N.S.\ H.) would like to thank the DAAD for financial
support during his stay in Bonn when part of the work was done.
\sn
We are indebted to the Max-Planck-Institut f\"ur Mathematik in
Bonn-Beuel for providing the computer time necessary for performing
the more complicated algebraic calculations reported in this paper.
\vfill
\eject
\chapsubtitle{Appendix A: A Proof of duality}
\mn
Duality has been proved in e.g.\ $\q{\elitzur} \q{\horn}$.
Still, we would like to present a slightly different approach
in this appendix. We derive duality by comparing the representation
$\rt$ (\secB.3) to the following representation $r$ that is usually
considered:
$$\eqalign{
r(\si_j) \state{i_1 \ldots i_j \ldots i_N}
            &= \om^{i_j} \state{i_1 \ldots i_j \ldots i_N} \cr
r(\Ga_j) \state{i_1 \ldots i_j \ldots i_N}
            &= \state{i_1 \ldots (i_j+1 \mod n) \ldots i_N}. \cr
}         \eqno{(\rm \appA.1)}$$
Note that $\rt(\Ga_j) = r(\si_j)$ and $\rt(\si_j) = r(\Ga_j^{+})$ and
that the representations $\rt$ and $r$ are unitarily equivalent.
\medskip
We recall the statement of section {\secD} before proving it.
Let $\hat{H}_N^{(n)}(\la,R^{ht},\ab_k^{ht},\a_k^{ht})$ be the Hamiltonian
with suitable normalization for
high-temperature expansions ($\hat{H}(\la) := \la H(\la^{-1})$ )
and $H_N^{(n)}(\tilde{\la},R^{lt},\ab_k^{lt},\a_k^{lt})$
be the Hamiltonian (\secA.1) with  corresponding parameters.
Furthermore, abbreviate the eigenspace of $r(\hat{Q})$
to eigenvalue $\om^{Q^{ht}}$ by $\H^{Q^{ht}}$ and that of $\rt(\hat{Q})$
to eigenvalue $\om^{Q^{lt}}$ by $\tilde{\H}^{Q^{lt}}$. Then
$\hat{H}_N^{(n)}(\la,R^{ht},\ab_k^{ht},\a_k^{ht})$ restricted
to $\H^{Q^{ht}}$ and
$H_N^{(n)}(\tilde{\la},R^{lt},\ab_k^{lt},\a_k^{lt})$ restricted to
$\tilde{\H}^{Q^{lt}}$ have the same spectra if
$$\eqalign{
Q^{lt} &= R^{ht} \ , \qquad  R^{lt} = Q^{ht} \ , \cr
\ab_k^{lt} &= \a_k^{ht} \ , \qquad  \a_k^{lt} = \ab_k^{ht} \ , \qquad
\tilde{\la} =  \la. \cr
}  \eqno{(\rm \secD.1)}$$
For the proof we fix the state $\GS{Q^{lt}}$ to be the ground state
(\secC.1) in $\tilde{\H}^{Q^{lt}}$. Then the
following states are a basis for $\tilde{\H}^{Q^{lt}}$
\footnote{${}^{1})$}{these may be different from e.g.\ (\secC.2) and
(\secC.10)}:
$$\state{Q; i_2 \ldots i_N} := \rt(\si_2^{-i_2}) \ldots
 \rt(\si_N^{-i_N}) \GS{Q^{lt}}.              \eqno{(\rm \appA.2)}$$
Note that this implies:
$$\rt(\si_1) \state{Q; i_2 \ldots i_N} =
\om^{Q^{lt}} \state{Q; (i_2+1) \ldots (i_N+1)}.       \eqno{(\rm \appA.3)}$$
Now consider the following intertwining isomorphism $I$:
$$I \state{Q; i_2 \ldots i_N} := \state{(-i_2) (i_2 -i_3) \ldots
                                     (i_{N-1} - i_N) (i_N + R^{lt}) }.
  \eqno{(\rm \appA.4)}$$
Note that the map (\appA.4) maps the ground states in both phases onto
each other.
It is now straightforward to check using the basis (\appA.2) that
$$I \rt(\si_{(j+1 \mod N)}) = r(\Ga_j \Ga_{j+1}^{+}) I \ , \qquad
I \rt(\Ga_j \Ga_{j+1}^{+}) = r(\si_j) I \ .
  \eqno{(\rm \appA.5)}$$
The observation that $I$ is a unitary map and $r$ and $\rt$ are unitarily
equivalent in conjunction with (\appA.5) proves duality.
\vfill
\eject
\chapsubtitle{References}
\mn
\settabs\+&\phantom{---------}&\phantom{
------------------------------------------------------------------------------}
& \cr
\+ &$\q{\hkn}$ & S.\ Howes, L.P.\ Kadanoff, M.\ denNijs & \cr
\+ &           & {\it Quantum Model
                 for Commensurate-Incommensurate Transitions} & \cr
\+ &           & Nucl.\ Phys.\ {\bf B215} (1983) p.\ 169 & \cr
\+ & $\q{\gehri}$
               & G.v.\ Gehlen, V.\ Rittenberg, {\it $\Zed_n$-Symmetric
                 Quantum Chains with an Infinite Set} & \cr
\+ &           & {\it of Conserved Charges and
                 $\Zed_n$ Zero Modes},
                 Nucl.\ Phys.\ {\bf B257} (1985) p.\ 351 & \cr
\+ & $\q{\dogra}$
               & L.\ Dolan, M.\ Grady, {\it Conserved Charges from
                 Self-Duality} & \cr
\+ &           & Phys.\ Rev.\ {\bf D25} (1982) p.\ 1587 & \cr
\+ & \q{\daviesA}
               & B.\ Davies, {\it Onsager's Algebra and Superintegrability}&\cr
\+ &           & Jour.\ Phys.\ A: Math.\ Gen.\ {\bf 23} (1990) p.\ 2245 & \cr
\+ & \q{\daviesB}
               & B.\ Davies &\cr
\+ &           & {\it Onsager's Algebra and the Dolan-Grady
                 Condition in the non-Self-Dual Case} & \cr
\+ &           & J.\ Math.\ Phys\ {\bf 32} (1991) p.\ 2945 & \cr
\+ & \q{\onsager}
               & L.\ Onsager & \cr
\+ &           & {\it Crystal Statistics.\ I.\ A Two-Dimensional
                 Model with an Order-Disorder Transition} & \cr
\+ &           & Physical Review {\bf 65} (1944) p.\ 117 & \cr
\+ & $\q{\albertiniA}$
               & G.\ Albertini, B.M.\ McCoy, J.H.H.\ Perk,
                 {\it Commensurate-Incommensurate Transition} & \cr
\+ &           & {\it in the Ground
                  State of the Superintegrable Chiral Potts Model} & \cr
\+ &           & Phys.\ Lett.\ {\bf 135A} (1989) p.\ 159 & \cr
\+ & $\q{\albertiniB}$
               & G.\ Albertini, B.M.\ McCoy, J.H.H.\ Perk,
                 {\it Level Crossing Transitions and the Massless} & \cr
\+ &           & {\it Phases of the Superintegrable Chiral Potts Chain},
                  Phys.\ Lett.\ {\bf 139A} (1989) p.\ 204  & \cr
\+ & $\q{\yang}$
               & H.\ Au-Yang, B.M.\ McCoy, J.H.H.\ Perk, Sh.\ Tang, M.L.\ Yan
                                                                    & \cr
\+ &           & {\it Commuting Transfer Matrices in the Chiral Potts Models:}
                                                                    & \cr
\+ &           & {\it Solutions of Star-Triangle Equations with Genus $>1$}
                                                                    & \cr
\+ &           & Phys.\ Lett.\ {\bf 123A} (1987) p.\ 219  & \cr
\+ & $\q{\baxter}$
               & R.J.\ Baxter, {\it The Superintegrable Chiral Potts Model}
                                                                    & \cr
\+ &           & Phys.\ Lett.\ {\bf 133A} (1988) p.\ 185 & \cr
\+ & $\q{\scm}$& B.M.\ McCoy, {\it The Chiral Potts Model: from Physics to
                 Mathemityatics and back} & \cr
\+ &           & Special functions ICM 90, Satellite Conf.\ Proc., ed.\
                 M.\ Kashiwara and T.\ Miwa, &\cr
\+ &           & Springer (1991) p.\ 245 & \cr
\+ & $\q{\hela}$
               & M.\ Henkel, J.\ Lacki & \cr
\+ &           & {\it Integrable Chiral $\Zed_n$
                 Quantum Chains and a New Class of Trigonometric Sums} & \cr
\+ &           & Phys.\ Lett.\ {\bf 138A} (1989) p.\ 105 & \cr
\+ & $\q{\tang}$
               & G.\ Albertini, B.M.\ McCoy, J.H.H.\ Perk, S.\ Tang & \cr
\+ &           & {\it Excitation Spectrum and Order Parameter for the
                 Integrable $N$-State Chiral Potts Model} & \cr
\+ &           & Nucl.\ Phys.\ {\bf B314} (1989) p.\ 741 & \cr
\+ & $\q{\weA}$
               & G.v.\ Gehlen, A.\ Honecker, {\it Multi-Particle Structure in
                 the $\Zed_n$-Chiral Potts Models} & \cr
\+ &           & Jour.\ Phys.\ A: Math.\ Gen.\ {\bf 26} (1993) p.\ 1275 & \cr
\+ & $\q{\dkcoy}$
               & S.\ Dasmahapatra, R.\ Kedem, B.M.\ McCoy & \cr
\+ &           & {\it  Spectrum and Completeness of the 3-State
                   Superintegrable Chiral Potts Model} & \cr
\+ &           & preprint ITB-SB 9211 (1992) & \cr
\+ & $\q{\gehkra}$
               & G.v.\ Gehlen, T.W.\ Krallman, {\it Phase Structure of
                  the $\Zed_3$-Chiral Potts Model} & \cr
\+ &           & in preparation & \cr
\+ & $\q{\gehlenph}$
               & G.v.\ Gehlen,
                 {\it Phase Diagram and Two-Particle Structure
                          of the $\Zed_3$-Chiral Potts Model} & \cr
\+ &           & preprint BONN-HE-92-18 (1992), hep-th/9207022,
                 to be published in proceedings of & \cr
\+ &           & {\it Int. Symposium
                 on Advanced Topis of Quantum Physics}, Taiyuan, China (1992)
               & \cr
\+ & $\q{\alcaraz}$
               & F.C.\ Alcaraz, A.L.\ Santos, {\it Conservation Laws for
                  \Zed(N) Symmetric Quantum} & \cr
\+ &           & {\it Spin Models and Their Exact Ground State Energies} & \cr
\+ &           & Nucl.\ Phys.\ {\bf B275} (1986) p.\ 436 & \cr
\+ & $\q{\fateev}$
               & V.A.\ Fateev, A.B.\ Zamolodchikov & \cr
\+ &           & {\it Conformal Quantum Field Theory Models in Two
                    Dimensions Having} & \cr
\+ &           & {\it $\Zed_3$ Symmetry},
                    Nucl.\ Phys.\ {\bf B280} (1987) p.\ 644 & \cr
\+ & $\q{\lykyanov}$
               & V.A.\ Fateev, S.L.\ Lykyanov, {\it The Models of
                   Two-Dimensional Conformal} & \cr
\+ &           & {\it Quantum Field Theory with $\Zed_n$ Symmetry} & \cr
\+ &           & Int.\ Jour.\ of Mod.\ Phys.\ {\bf A3} (1988) p.\ 507 & \cr
\+ &$\q{\zam}$ & A.B.\ Zamolodchikov & \cr
\+ &           & {\it Infinite Additional Symmetries in Two-Dimensional
                   Conformal Quantum Field}  & \cr
\+ &           & {\it Theory}, Theor.\ Math.\ Phys.\ 65 (1986) p.\ 1205  & \cr
\+ & $\q{\mccoyadv}$
               & G.\ Albertini, B.M.\ McCoy, J.H.H.\ Perk, {\it Eigenvalue
                    Spectrum of the} & \cr
\+ &           & {\it Superintegrable Chiral Potts Model},
                    Adv.\ Studies in Pure Math.\ {\bf 19} (1989) p.\ 1 & \cr
\+ & $\q{\chrihen}$
               & Ph.\ Christe, M.\ Henkel & \cr
\+ &           & {\it Introduction to Conformal
                 Invariance and its Applications to Critical Phenomena} & \cr
\+ &           & preprint UGVA/DPT 1992/11 - 794, to appear in
                 Springer Lecture Notes in Physics & \cr
\+ & $\q{\gehritell}$
               & G.v.\ Gehlen, V.\ Rittenberg & \cr
\+ &           & {\it The Ashkin-Teller Quantum Chain and
                      Conformal Invariance} & \cr
\+ &           & Jour.\ Phys.\ A: Math.\ Gen.\ {\bf 20} (1987) p.\ 227 & \cr
\+ & $\q{\baym}$
               & G.\ Baym, {\it Lectures on Quantum Mechanics} & \cr
\+ &           & Benjamin/Cummings Publishing (1969), 3rd printing (1974),
                 chapter 11 & \cr
\+ & $\q{\hoeger}$
               & C.\ Hoeger, G.v.\ Gehlen, V.\ Rittenberg & \cr
\+ &           & {\it Finite-Size Scaling for Quantum Chains with an
                      Oscillatory Energy Gap} & \cr
\+ &           & Jour.\ Phys.\ A: Math.\ Gen.\ {\bf 18} (1985) p.\ 1813 & \cr
\+ & $\q{\weB}$
               & G.v.\ Gehlen, A.\ Honecker & \cr
\+ &           & {\it Quasiparticle Spectrum and Correlation
                 Functions of the Chiral Potts Model} & \cr
\+ &           & in preparation & \cr
\+ & $\q{\elitzur}$
               & S.\ Elitzur, R.B.\ Pearson, J.\ Shigemitsu & \cr
\+ &           & {\it Phase Structure of Discrete Abelian Spin and
                  Gauge Systems} & \cr
\+ &           & Phys.\ Rev.\ {\bf D19} (1979) p.\ 3698 & \cr
\+ & $\q{\horn}$
               & D.\ Horn, M.\ Weinstein, S.\ Yankielowicz & \cr
\+ &           & {\it Hamiltonian Approach to \Zed(N) Lattice
                      Gauge Theories} & \cr
\+ &           & Phys.\ Rev.\ {\bf D19} (1979) p.\ 3715 & \cr
\+ & $\q{\albcoy}$
               & G.\ Albertini, B.M.\ McCoy, {\it Correlation Functions
                 of the Chiral Potts Chain} & \cr
\+ &           & {\it from Conformal Field Theory
                 and Finite-Size Corrections} & \cr
\+ &           & Nucl.\ Phys.\ {\bf B350} (1990) p.\ 745 & \cr
\vfill
\end